\documentclass[showpacs,prd,twocolumn]{revtex4}
\usepackage{pifont}
\usepackage{bbding}
\usepackage{mathrsfs}
\usepackage{longtable,lscape} \usepackage{multirow}
\usepackage{txfonts,bm}
\usepackage{amssymb}
\usepackage{indentfirst}
\usepackage{graphicx,booktabs}
\usepackage{color}
\usepackage{hyperref}

\begin{document}

\title{Perfect $DD^*$ molecular prediction matching the $T_{cc}$ observation at LHCb
}
\author{Ning Li$^{1}$}\email{lining59@mail.sysu.edu.cn}
\author{Zhi-Feng Sun$^{2, 3,4}$}\email{sunzf@lzu.edu.cn}
\author{Xiang Liu$^{2,3,4}$}\email{xiangliu@lzu.edu.cn}
\author{Shi-Lin Zhu$^{5}$}\email{zhusl@pku.edu.cn}
\affiliation{$^1$School of Physics, Sun Yat-Sen University, Guangzhou 510275, China\\
$^2$School of Physical Science and Technology, Lanzhou University, Lanzhou 730000, China\\
$^3$Research Center for Hadron and CSR Physics, Lanzhou University and Institute of Modern Physics of CAS, Lanzhou 730000, China \\
$^4$ Lanzhou Center for Theoretical Physics, Lanzhou University, Lanzhou, Gansu 730000, China \\
$^5$ School of Physics and Center of High Energy Physics, Peking
University, Beijing 100871, China }
\date{\today}

\begin{abstract}

In 2012, we investigated the possible molecular states composed of
two charmed mesons [Phys.Rev. {\bf D 88}, 114008 (2013), arXiv:1211.5007 [hep-ph](2012)] .
The $D^*D$ system with the quantum numbers of $I(J^P)=0(1^+)$ was
found to be a good candidate of the loosely bound molecular state.
This state is very close to the $D^*D$ threshold with a binding
energy around 0.47 MeV. This prediction was confirmed by the new
LHCb observation of $T_{cc}^+$ [see Franz Muheim's talk at the
European Physical Society conference on high energy physics 2021].

\end{abstract}

\maketitle

%%%%%%%%%%%%%%%%%%%%%%%%%%%%%%%%%%%%%%%%%%%%%%%%%%%
%\section{Introduction} \label{Introduction}
%%%%%%%%%%%%%%%%%%%%%%%%%%%%%%%%%%%%%%%%%%%%%%%%%%%

Since the Belle Collaboration discovered the $X(3872)$ in 2003
\cite{Choi:2003ue}, a series of charmoniumlike $XYZ$ states which
are close to the thresholds of two hadrons have been observed. There
exists difficulty to describe these states in the traditional hadron
picture, which has posed a big challenge and brought valuable
opportunities to the whole community. The study of these new exotic
hadron states has become the hot new frontier of the hadron physics.

Especially, in the last three years, the LHCb Collaboration has been
playing a key role in finding new hadronic states. In 2019, the
characteristic mass spectrum of three $P_c$ states, the
$P_c(4312)^+$, $P_c(4440)^+$ and $P_c(4457)^+$\cite{Aaij:2019vzc}
provide strong evidence of the existence of the hidden-charm
molecular pentaquark. Additionally, LHCb reported the evidence of
the $P_{cs}(4459)$ \cite{LHCb:2020jpq} in the $J/\psi\Lambda$
invariant mass distribution. And then, several hidden-charm
tetraquark candiadates (the $Z_{cs}(4000)^+$ and $Z_{cs}(4220)^+$
\cite{LHCb:2021uow}) and possible single-charm tetraquarks, the
$d\bar{c}u\bar{s}$ tetraquarks ($X_0(2900)$ and $X_1(2900)$
\cite{LHCb:2020bls, LHCb:2020pxc}) were observed. All these
measurements have caught much attention from both theorists and
experimentalists~\cite{CPL:Wu,CPL:Chen}.

Although so many multiquark state candidates have been reported in
experiment, it is not the end of whole story. At the European
Physical Society conference on high energy physics 2021, Franz
Muheim gave a talk on the ``LHCb Highlights", which shows that LHCb
observed the $T_{cc}^+$ by analyzing the $D^0D^0\pi^+$ invariant
mass spectrum~\cite{LHCb:2021vvq}. The difference between  mass of this
state  and the $D^{*+}D^0$ threshold is $-273 \pm 61 \pm 5
^{+11}_{-14}$ keV, and its width is $410\pm 165 \pm 43 ^{+18}_{-38}$
keV \cite{LHCb:2021vvq}. Obviously, as the first observation of the
double-charm tetraquark, the $T_{cc}$ has the $cc\bar u\bar d$
configuration. In this note we shall briefly introduce our previous
work on this topic~\cite{Tcc}. An extensive review of the $T_{cc}$
system can be found in Ref. \cite{Liu:2019zoy}.

%
%%%%%%%%%%%%%%%%%%%%%%%%%%%%%%%%%%%%%
%\section{Numerical Results}\label{Results}
%

In 2012, we investigated the system of the $D^{(*)}D^{(*)}[I(J^P) =
0(1^+)]$ very carefully by taking into account  the coupled channel
effect and the $S$-$D$ mixing effect, which play an important role
in the formation of the loosely bound deuteron \cite{CDBonn}. With
the one-pion-exchange force only, we found a loosely
$D^{*}D[I(J^P)=0(1^+)]$ bound state with the binding energy $1.24$
MeV for a reasonable cutoff $1.05$ GeV which is introduced to
suppress the very high-momentum or short-range contribution. The
corresponding root-mean-square radius is $3.11$~fm which is
comparable to the size of the well-known deuteron (about $2.0$~fm).
There exists a small contribution around $2.79\%$ from the
$D^*D^*(^3S_1)$ channel because of the large mass gap (about
$140$ MeV) between the thresholds of $DD^*$ and $D^*D^*$. However,
the probability of the $D$-wave interaction is tiny, $0.73\%$ for
the $DD^*({}^3D_1)$ and $0.08\%$ for the $D^*D^*(^3D_1)$, which
indicates that the $D^{*}D[I(J^P) = 0(1^+)]$ molecular state is
almost dominated by the $S$-wave interaction. 
We should mention that for the pion exchange in the crossed
channel, the potential is complex because the exchanged pion is
on-shell due to the large mass difference between $D$ and $D^*$.
With the same approach in the study of $X(3872)$, we keep the real
part of the potential. Such a formalism has no effects on the spectrum 
of the $DD^*$ system.

When the heavier $\rho$ and $\omega$ exchanges  are included, the
binding becomes deeper for the same cutoff as that with the
one-pion-exchange potential only. With a smaller cutoff $0.95$~GeV,
there exists a loosely $D^{(*)}D^{(*)}[I(J^P) = 0(1^+)]$ state with
a binding energy $0.47$~MeV and hence with the mass $3875.38$~MeV
which can be identified as the newly observed $T_{cc}$ by the LHCb
Collaboration. In such a case, the heavier $\rho$ and $\omega$
exchanges cancel each other significantly since for the isospin-zero
system the isospin factor of the $\rho$ meson exchange is $-3$ while
that of the $\omega$ meson exchange is $1$, and the residual force
is helpful to strengthen the binding. Additionally, the
contributions from the $\eta$ and $\sigma$ exchanges are very small.
The $T_{cc}$ and $X(3872)$ shares the same one-pion-exchange
potential. Their long-range dynamics is similar and correlated to
each other. If the X(3872) is a loosely bound molecular state, the
existence of the $X(3872)$ implies the existence of the $T_{cc}$.
Numerically, the binding energy of the $T_{cc}$ depends weakly on
the cutoff, which is similar to the case of the $X(3872)$. For
example, the binding energy of the $T_{cc}$ is $0.47$~MeV for the
cutoff $= 0.95$~GeV and $18.72$~MeV for cutoff $= 1.05$~GeV.

\begin{figure}
  \centering
  \includegraphics[width=0.48\textwidth]{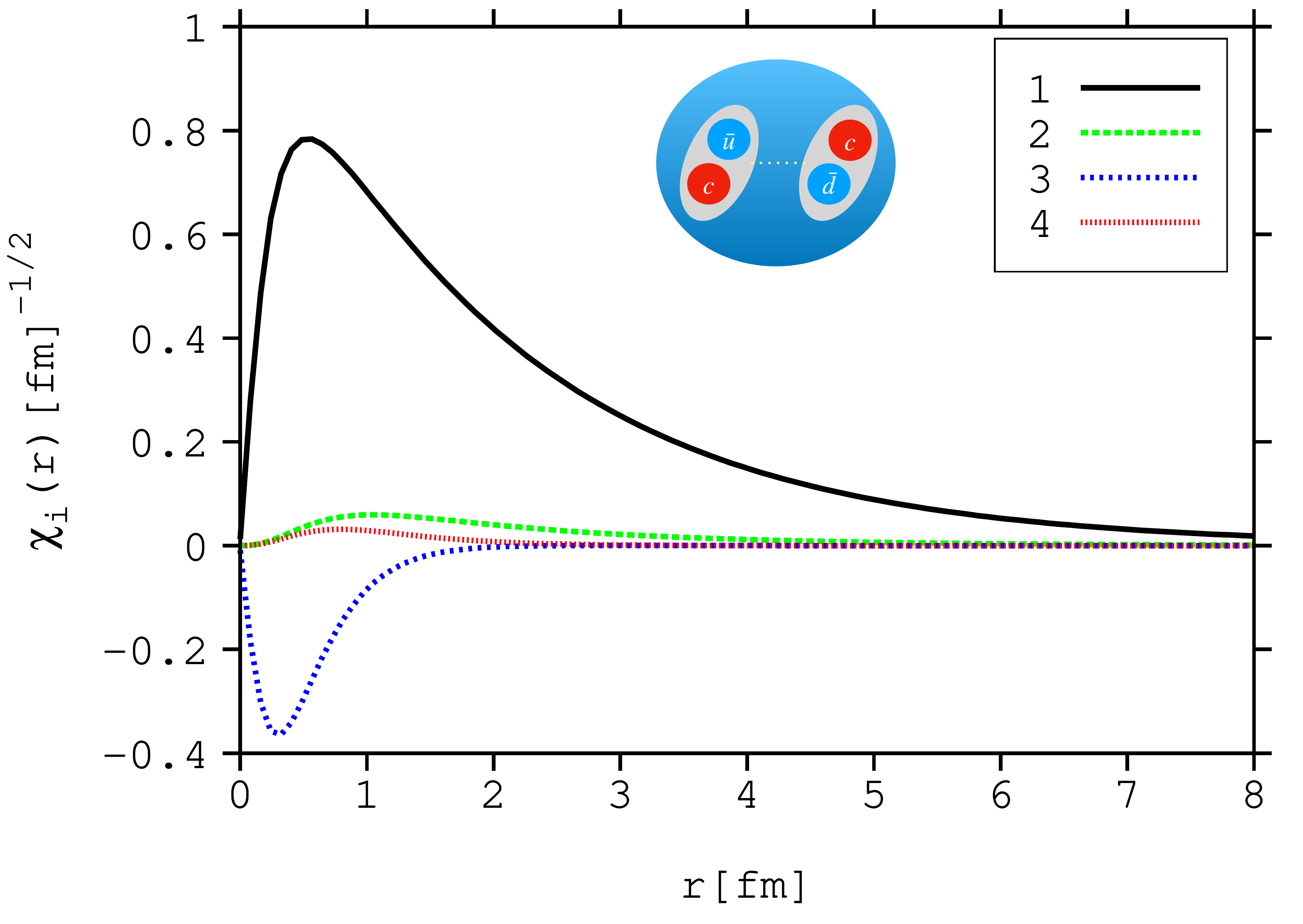}
  \caption{(Color online). The wave function ($\chi(r) = r\psi(r)$) of the $D^{(*)}D^{(*)}I(J^P)=0(1^+)$. ``1" denotes
  the channel $[DD^*(^3S_1)]$, ``2" for $[DD^*(^3D_1)]$, ``3" for $[D^*D^*(^3S_1)]$ and ``4" for $D^*D^*(^3D_1)]$}\label{plot:wfunction}
\end{figure}

The effects of the $\sigma$, $\rho$ and $\omega$ exchanges are also
analyzed by turning off the contribution of the $\pi$ and $\eta$
exchanges. A loosely bound state with a binding energy $0.78$~MeV
and root-mean-square radius $3.74$~fm was obtained when the cutoff
parameter is tuned to be $1.44$ GeV which is larger than $1.05$ GeV
used for the one-pion-exchange case. The contribution of the
long-range pion exchange is larger than that of the heavier vector
meson exchange in the formation of the loosely bound
$D^{(*)}D^{(*)}[I(J^P)=0(1+)]$ state. The wave function is shown in
Fig.~\ref{plot:wfunction} from which one can see that there is no
node except the origin. In other words, it is really a ground state.

In summary, in 2012 we predicted a loosely bound state of the
$DD^*[I(J^P) = 0(1^+)]$ which can be perfectly identified as the
newly observed $T_{cc}$ by the LHCb Collaboration. For this state,
the probability of the $D$-wave interaction is very small, and it
arises from the $S$-wave interaction. There is small contribution
from the $D^*D^*$ channel. The long-range pion exchange is strong
enough to form the loosely bound state, and the medium-range $\eta$
and $\sigma$ exchanges and the short-range $\rho$ and $\omega$
exchanges are helpful to strengthen the binding.

In Ref.~\cite{Molina:2010tx}, the authors also studied the doubly
charmed systems within the hidden gauge formalism in a
coupled-channel unitary approach. For the $D^*D^*$ system with $C=2,
S=0$ and $I=0$, they only obtained a bound state with quantum number
$I(J^P)=0(1^+)$. However, the pole appeared at 3969 MeV, which is
about 100 MeV larger than our result. In contrast, we considered the
coupled channel effect between the $DD^*$ and $D^*D^*$. Actually,
what we obtained is a $DD^*$ bound state instead of a $D^*D^*$ bound
state.

\begin{figure}[hbpt]
\centering
\includegraphics[width=0.48\textwidth]{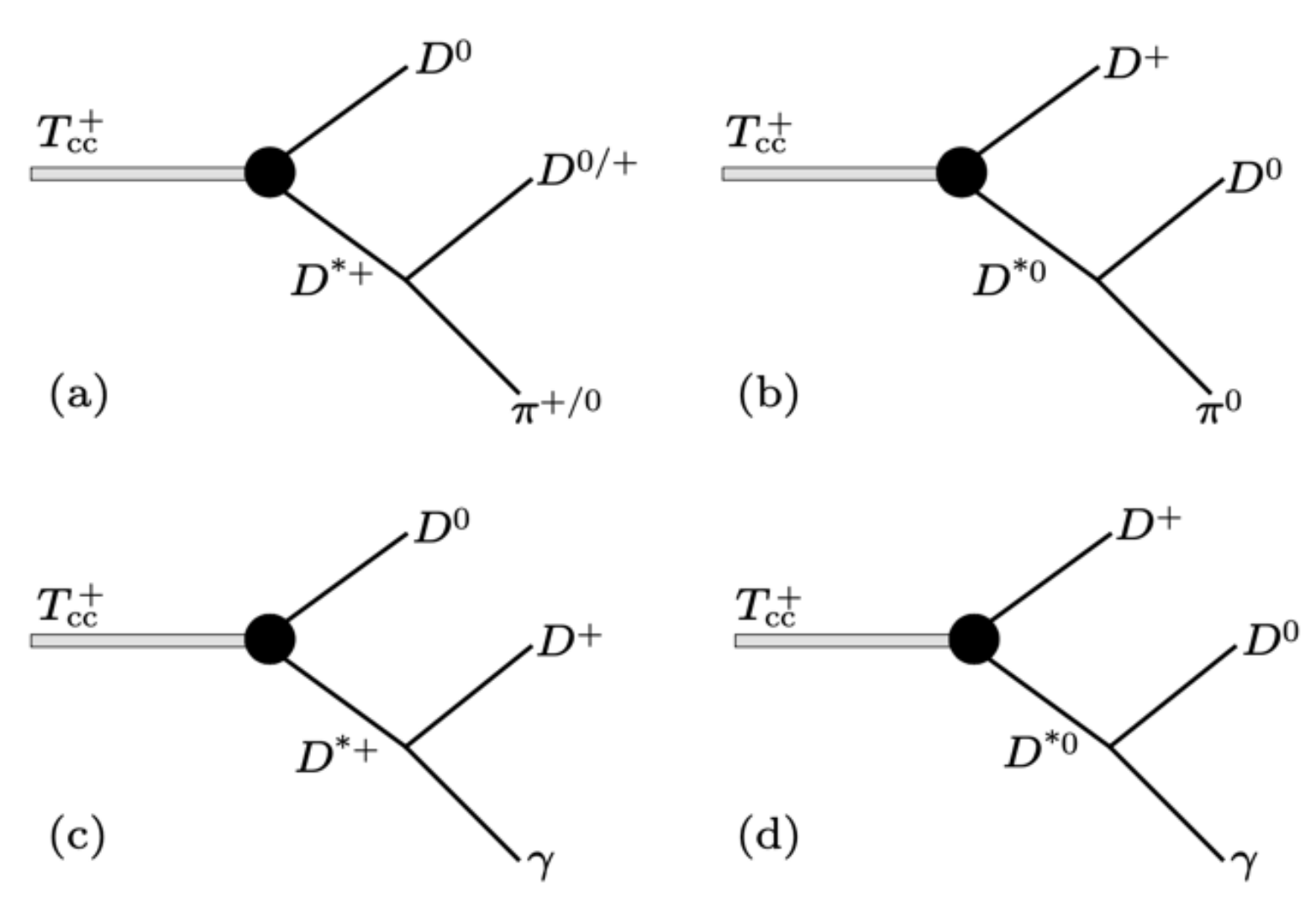}
\caption{The allowed strong and radiative decays of
$T_{cc}$.\label{decay}}
\end{figure}

The state $DD^*[0(1^+)]$ cannot decay into a double charm baryon
plus a light baryon. The masses of the lightest doubly-charmed
baryon and light baryon are 3518 MeV and 938 MeV, respectively,
corresponding to $\Xi_{cc}^+$ and proton as listed in Particle Data
Group \cite{0954-3899-37-7A-075021}. The mass of the molecular state
is around 3875 MeV and much smaller than the sum of the masses of a
doubly-charmed baryon and a light baryon. Therefore such a decay is
kinematically forbidden. However, the heavy vector meson $D^*$
within the exotic molecular state mainly decays to $D\pi$ via the
strong interaction as shown in Fig.~\ref{decay} (a) and (b). It also
decays into $D\gamma$ which is illustrated in Fig.~\ref{decay} (c)
and (d). The main decay modes should be $DD\gamma$ and $DD\pi$, and
the $D^{(*)}$ meson may also decay via the weak interaction.

The above typical decay modes provide important information to
experimental investigation of the properties of $T_{cc}$. With
further theoretical and experimental progress, we shall gain new
insights into this structure. The observation of this doubly charmed
structure has opened a new window for the exotic hadron states in
this exciting era, which is beyond the hidden charm "exotic" states
which have been widely investigated since 2003 when the $X(3872)$
was first observed by the Belle Collaboration.

\section*{ACKNOWLEDGMENTS}
This work is supported by the China National Funds for Distinguished
Young Scientists under Grant No. 11825503, National Key Research and
Development Program of China under Contract No. 2020YFA0406400, and
the 111 Project under Grant No. B20063, the National Natural Science
Foundation of China. This project is also supported by the National
Natural Science Foundation of China under Grant No. 11975033, No.
12070131001 and  No. 12047501. \vfil

\end{document}